\documentclass[prl,twocolumn,a4paper,superscriptaddress,showpacs]{revtex4}
\usepackage{graphicx}
\usepackage{amsmath}
\usepackage{amsfonts}
\usepackage{amssymb}
\newcommand{\be}{\begin{equation}}
\newcommand{\ee}{  \end{equation}}
\newcommand{\ba}{\begin{eqnarray}}
\newcommand{\ea}{  \end{eqnarray}}

\begin{document}

\title{Mono-parametric quantum charge pumping: interplay between spatial interference
and photon-assisted tunneling}
\author{Luis E. F. Foa Torres}
 \email{luisfoa@gmail.com}
 \altaffiliation{Present address: DRFMC/SPSMS/GT, CEA-Grenoble, 17 rue des Martyrs, 38054 Grenoble cedex 9, France.}
\affiliation{International Center for Theoretical Physics, Strada Costiera
11, 34014 Trieste, Italy}
\date{September 5, 2005}
\begin{abstract}
We analyze quantum charge pumping in an open ring with a dot embedded in one
of its arms. We show that cyclic driving of the dot levels by a
\textit{single} parameter leads to a pumped current when a static magnetic flux is
simultaneously applied to the ring. Based on the computation of the
Floquet-Green's functions, we show that for low driving frequencies $\omega_0$,
the interplay between the spatial interference through the ring plus
photon-assisted tunneling gives an average direct current (dc) which is
proportional to $\omega_0^{2}$. The direction of the pumped current can be reversed by changing the applied magnetic field.
\end{abstract}
\maketitle

%
%
%
%
%

%
%
%
%
%
%

\section{Introduction}

A directed current (dc) is usually associated to a dissipative flow of
the electrons in response to an applied bias voltage. However, in systems of
mesoscopic scale a dc current can be generated even at \textit{zero} bias. This
captivating quantum coherent effect is called \textit{quantum charge pumping
}\cite{Thouless,Spivak1995,Brouwer PRB1998} and it is of considerable interest
both theoretically \cite{Thouless,Spivak1995,Brouwer PRB1998,Stafford and
Wingreen PRL1996,ShutenkoPRB2000,Moskalets and Buttiker PRB 2002 FST QP,Kim
FST QP,Caio} and experimentally \cite{Kouwenhoven,Switkes Science1999}. A
device capable of providing such effect is called a \textit{quantum pump} and
typically involves the cyclic change of two device-control parameters with a
frequency $\omega_{0}$. The operational regime of the pump can be
characterized according to the relative magnitude between $\omega_{0}$ and the
inverse of the time taken for an electron to traverse the sample, $1/\tau_{T}%
$. When $\omega_{0}\ll1/\tau_{T}$ the pump is in the so called
\ \textit{``adiabatic}'' regime, whereas the opposite case, $\omega_{0}%
\gg1/\tau_{T}$, the pump is in the \textit{``non-adiabatic''} regime.

%
%
%
%
%
%
For adiabatic pumping, Brouwer \cite{Brouwer PRB1998} gave an appealing
approach which is based on a scattering matrix formulation to low frequency ac
transport due to B\"{u}ttiker, Thomas and Pr\^{e}tre \cite{Buttiker Thomas
Pretre}. In this formulation, the pumped current which flows in response to a
the cyclic variation of a set $\left\{  X_{j}\right\}  $ of device-control
parameters is expressed in terms of the scattering matrix $S(\left\{
X_{j}\right\}  )$ of the system. One of the outcomes of this parametric
pumping theory, which is valid in the low frequency regime ($\omega_{0}%
\ll1/\tau_{T}$) and up to first order in frequency, is that the charge pumped
during a cycle is proportional to the area enclosed by the path in the
scattering matrix parameter space. Thus, to have a non-vanishing pumped charge
at least \textit{two} time-dependent parameters that oscillate with a
frequency $\omega_{0}$ and with a non-vanishing phase difference $\varphi$
between them are needed.

\begin{figure}[ptb]
%
%
%
%
%
\includegraphics[width=6.0cm]{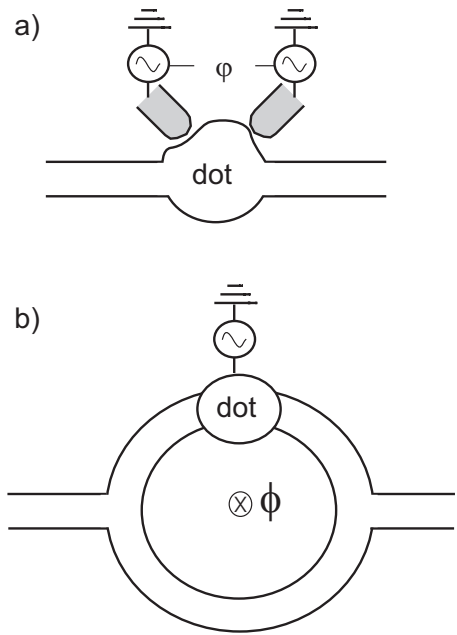} \vspace{0.5cm}\caption{a) Schematic
representation of a typical quantum pump consisting in an open dot driven by
two out of phase time-dependent gate voltages. b) Scheme of the system
considered in this work, a ring connected to two leads. The ring, which is
threaded by a magnetic flux, contains a dot embeded in one of its arms. Charge
pumping is obtained by driving the dot levels through a time-periodic
potential. }%
\label{fig-schemes}%
\end{figure}

%
%
%
%
%
%
%
%
In this context, a natural question that arises is whether a pumped current
can be obtained using a \textit{single} time-dependent parameter. In most of
the works considered up to now, at least \textit{two} parameters are used to
obtain pumping. A typical configuration that has been extensively studied
theoretically and experimentally \cite{Switkes Science1999} consists of a dot
connected to two leads with two out of phase time-dependent gate voltages that
produce cyclic changes in its shape (see Fig. \ref{fig-schemes}a). In
contrast, pumps based on a \textit{single} parameter variation have attracted
much less attention. This is partly due to the fact that no pumping can be
obtained from them in the low frequency regime up to first order in
$\omega_{0}$. Hence, obtaining a non-trivial result requires to go beyond the
adiabatic limit described by the standard parametric pumping theory
\cite{Brouwer PRB1998} as in Refs. \cite{ShutenkoPRB2000,FalkoSovPhys,VavilovPhaseSpacePicture}. In spite of giving a current which at low frequencies is \textit{a priori} weaker than the one
obtained using a two-parameter variation, they can give comparable pumped
currents at intermediate and high frequencies \cite{note on rectification}.
Besides, the understanding of such ``mono-parametric pumps'' constitute a
necessary step in the comprehension of driven systems.

Previous theoretical studies in this direction include the works by Kravtsov,
Yudson and Aronov \cite{KravtsovPRL1993,Aronov and Kravtsov 1993} where
pumping in a ring (not connected to leads) threaded by a time-dependent flux
was studied. In Ref. \cite{Guo finite frequency PRB 2002} Guo and
collaborators considered the case in which the height of one of the barriers
of a double barrier system connected to external leads is modulated
periodically. This modulation \textit{dynamically} breaks the inversion
symmetry of the system producing a pumped current. Other theoretical works aiming at the frequency dependence of the pumped current for situations beyond the adiabatic approximation were also reported in Refs. \cite{FalkoSovPhys,VavilovPhaseSpacePicture}. On the other hand,
experimental studies using a single parameter modulation have been reported
recently \cite{DiCarloMarcusPRL2003,VavilovMarcus PRB2005} showing predominant
rectification effects \cite{Brouwer-rectification} at low frequencies and
quantum pumping in the high frequency regime.

In this work we focus in mono-parametric pumping in systems connected to
external leads. Specifically, we study a quantum pump consisting of a ring
with a dot which is subjected to a \textit{single} time-periodic gate voltage
embedded in one of its arms as represented in Fig.\ref{fig-schemes}b
\cite{noteRingPlus2parametersButtiker}. The ring is threaded by a
\textit{static} magnetic field which produces the left-right symmetry breaking
needed for pumping. The unique role of the time-dependent parameter in this
pump is to provide for photon-assisted channels. We show that in the low
frequency regime ($\omega_{0}\ll1/\tau_{T}$) as a result of the interplay
between spatial interference through the ring and photon-assisted processes,
this device produces a dc current which is proportional to $\omega_{0}^{2}$ and whose direction can be reversed by tuning the applied magnetic field.

Our theoretical framework mostly follows Ref. \cite{CamaletPRL2003} and is
based on the use of Floquet's theory \cite{Shirley,Sambe} to write the average
current in terms of the Fourier components of the retarded Green's functions
for the system. However, instead of solving an eigenvalue problem as in Refs.
\cite{CamaletPRL2003,Kohler PhysRep}, we completely rely on the computation of
the Floquet-Green's functions for the system. The resulting picture is that of
an equivalent time-independent problem in a higher dimensional space (similar
to the one previously obtained for electron-phonon interactions
\cite{AndaMaklerPastawskiBJP1994,Bonca1995}) and is specially suited for
discrete Hamiltonians offering thus a promising application to molecular
systems \cite{PecchiaRepProgPhys2004}.

This work is organized as follows. First we briefly introduce our theoretical
tools and present the case of a driven double barrier as an example to
motivate the subsequent discussion of mono-parametric quantum pumping. Then we
focus in our model system, discussing our analytical and numerical results.

\section{Theory}

In this section we introduce the theoretical tools that will be used to
address our specific problem. In order to keep the discussion general we
consider at this point a generic system consisting of a sample region that is
connected to two leads (left and right). The time-dependent Hamiltonian can be
written in the form:%
\[
H=H_{sample}(t)+H_{leads}+H_{contacts},
\]
where the terms correspond to the contributions from the sample, the leads and
the sample-leads coupling respectively. Following Ref. \cite{CamaletPRL2003}
we focus on the regime of quantum coherent transport and electron-electron
interactions are not considered.

We consider a situation in which both leads are in thermal equilibrium with a
common chemical potential, i.e., $f_{L}(\varepsilon)=f_{R}(\varepsilon)\equiv
f(\varepsilon)\,$for all energies. Then, the current averaged over one period
of the modulation is given by \cite{CamaletPRL2003,Kohler PhysRep}:%

\begin{equation}
\overline{I}=\frac{e}{h}\sum_{n}\int d\varepsilon\left[  T_{R\leftarrow
L}^{(n)}(\varepsilon)-T_{L\leftarrow R}^{(n)}(\varepsilon)\right]
f(\varepsilon) \label{eq-averageCurrent}%
\end{equation}
where
\begin{equation}
T_{R\leftarrow L}^{(n)}(\varepsilon)=2\Gamma_{R}(\varepsilon+n\hbar\omega
_{0})\left|  G_{RL}^{(n)}(\varepsilon)\right|  ^{2}\,2\Gamma_{L}(\varepsilon)
\label{eq-T_n}%
\end{equation}
are the transmission probabilities for an electron with energy $\varepsilon$
from left to right involving the absorption (or emission) of an energy
$n\hbar\omega_{0}\,$(and similarly for $T_{L\leftarrow R}^{(n)}(\varepsilon
)$). Here $G^{(n)}(\varepsilon)=\frac{1}{T}\int_{0}^{T}dt\,e^{-in\omega_{0}%
t}G(t,\varepsilon)$ are the coefficients of the Fourier decomposition of the
retarded Green's function $G^{R}(t,\varepsilon)$, and $\Gamma_{L(R)}%
(\varepsilon)$ are given by the imaginary part of the retarded self-energy
correction due to the corresponding lead, $\Gamma_{L(R)}(\varepsilon
)=-\operatorname{Im}\Sigma_{L(R)}(\varepsilon)$.

Equation (\ref{eq-averageCurrent}) was derived in Ref. \cite{CamaletPRL2003}
by solving the Heisenberg equations of motion for the creation and
annihilation operators and taking advantage of the time-periodicity of the
Hamiltonian. The transmission probabilities $T_{\alpha\leftarrow\beta}%
^{(n)}(\varepsilon)$ were expressed in terms of the Fourier components
$G_{\beta\alpha}^{(n)}(\varepsilon)$ of the retarded Green's functions. This
procedure is, for non-interacting electrons, formally equivalent to the use of
the Keldysh formalism \cite{GLBE2,JauhoWingreenMeir1994} but in contrast with
previous works along that path, the time-periodicity of the Hamiltonian is
exploited through the use of Floquet's theory \cite{Intro-Floquet}.

In Ref. \cite{CamaletPRL2003} the Fourier components $G_{\beta\alpha}%
^{(n)}(\varepsilon)$ of the retarded Green's functions were written (after
tracing over the degrees of freedom in the leads) in terms of the solutions of
Floquet's equation \cite{Shirley,Sambe} for the sample region. While the
Floquet's states and their corresponding quasi-energies can be obtained
numerically, this can take a significant computational power depending on the
system size. Here, we use instead a different strategy: the essential idea is
to write Eq. (\ref{eq-averageCurrent}) completely in terms of the
Floquet-Green's functions.

To such end we note that the Fourier coefficients $G^{(n)}(\varepsilon)$ can
be written as (see appendix):%
\begin{equation}
G_{\alpha\beta}^{(n)}(\varepsilon)=G_{(\alpha,n),(\beta,0)}^{F}(\varepsilon),
\label{eq-FourierComponents-FloquetGF}%
\end{equation}
where the Floquet-Green's functions
\begin{equation}
G_{(\alpha,n),(\beta,0)}^{F}\equiv\left\langle \alpha,n\right|  (\varepsilon
I-H_{F})^{-1}\left|  \beta,0\right\rangle
\end{equation}
are defined in terms of the Floquet Hamiltonian:
\begin{equation}
H_{F}=H(t)-i\hbar\frac{\partial}{\partial t}. \label{eq-Floquet Hamiltonian}%
\end{equation}
Notice that both $H_{F}$ and $G^{F}$ are defined in the composed Hilbert space
$R\otimes T$, where $R$ is the space of functions in real space and $T$ is the
space of periodic functions with period $\tau=2\pi/\omega_{0}$.$\ $The space
$T$ is spanned by the set of orthonormal Fourier vectors $\left\langle
t\right.  \left|  n\right\rangle \equiv\exp(in\omega_{0}t),$ where $n$ is an
integer. A suitable basis for this so called \textit{Floquet or Sambe space
}\cite{Sambe}, $R\otimes T$ , is thus given by $\{\left|  i,n\right\rangle
\equiv\left|  i\right\rangle \otimes\left|  n\right\rangle \},$ where $\left|
i\right\rangle $ corresponds to a state localized at site $i$.

Substituting Eq. (\ref{eq-FourierComponents-FloquetGF}) in Eqs. (\ref{eq-T_n})
and (\ref{eq-averageCurrent}) we write the average current completely in terms
of the Floquet-Green's functions $G^{F}$ (see appendix). The key point here is
that this renders an equivalent time-independent problem in a higher
dimensional space, $R\otimes T$. Therefore, the full power of the recursive
Green's functions techniques \cite{recursive Greens functions} can be used accordingly.

To compute these functions we consider only the Floquet states $\left|
j,n\right\rangle $ within some range for $n$, i.e., $\left|  n\right|  \leq
N_{\max}$. This range can be successively expanded until the answer converges,
giving thus a variational (non-perturbative) method. The resulting scheme is
similar to the one introduced in
Refs.\cite{AndaMaklerPastawskiBJP1994,Bonca1995} for the problem of
phonon-assisted electron transport \cite{ChemPhyse-ph}. The main differences
\cite{DFMartinez} are that: \textit{a)} for phonons the temperature enters
naturally in the population of the different channels, \textit{b)} that the
phonon spectrum is bounded from below and \textit{c)}, that the matrix element
for phonon emission and absorption depend on the number $N$ of phonons present
in the system before the scattering process.
%
%
%
%
%
%
%
%
%
%
%
%
%
%
%
%
%
%
%
%
%
%
%
%
%
%
%
%
%
%

In order to fix ideas and to motivate the subsequent discussion, let us see
how this picture works for the case of a driven double barrier and then we
turn into the study of our model system.

\subsection{Example: Driven double barrier}

\begin{figure}[ptb]
%
%
%
%
%
\includegraphics[width=7.0cm]{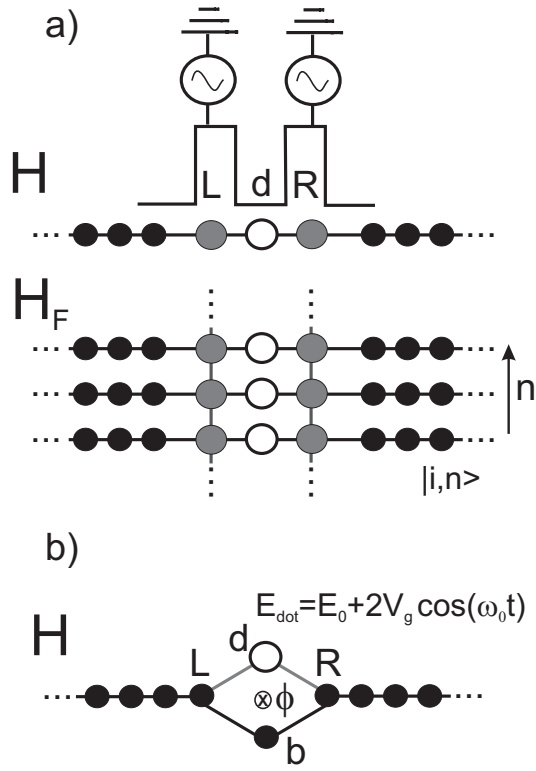} \vspace{0.5cm}\caption{a) Top: Scheme
of a double barrier system driven by time-dependent gate voltages applied to
each of the barriers. Bottom: Representation of the Floquet Hamiltonian
corresponding to the tight-binding model introduced in the text for the system
shown in the top. Circles correspond to different states $\left|
i,n\right\rangle $ in Floquet space and lines are off-diagonal matrix
elements. States along a vertical line correspond to the same spatial states
with different number $n$ of \ photon quanta. Note that the phase of the
vertical matrix elements connecting different Floquet states generally depend
on the direction of the transition. For example, the matrix element of the
Floquet Hamiltonian connecting the $\left|  L,n\right\rangle $ state with
$\left|  L,n+1\right\rangle $ is $V_{gL}\exp(i\varphi_{L})$ while the matrix
element for the reverse process is $V_{gL}\exp(-i\varphi_{L})$. b) Representation of the model Hamiltonian used in the text for the situation in Fig. 1b. }%
\label{fig-models}%
\end{figure}

Let us consider a system as depicted in the top of Fig. \ref{fig-models}a. A
simple Hamiltonian for that situation is given by:%
\begin{align*}
H_{sample}  &  =\widetilde{E_{L}}(t)\,c_{L}^{+}c_{L}^{{}}+E_{0}\,c_{d}%
^{+}c_{d}^{{}}+\widetilde{E_{R}}(t)\,c_{R}^{+}c_{R}^{{}}+\\
&  +V_{L,d}(c_{L}^{+}c_{d}^{{}}+h.c.)+V_{R,d}(c_{R}^{+}c_{d}^{{}}+h.c.).
\end{align*}
where $c_{\alpha}^{+}$ ($c_{\alpha}^{{}}$) are the creation (destruction)
operators at site $\alpha$, $\widetilde{E_{L}}=E_{L}+2V_{gL}\cos(\omega
_{0}t+\varphi_{L})$ and $\widetilde{E_{R}}=E_{R}+2V_{gR}\cos(\omega
_{0}t+\varphi_{R})$ are the energies of the barrier sites which are modulated
periodically. Only one energy state $E_{0}$ inside the double barrier is
considered. The leads are regarded as one-dimensional tight-binding chains
which are coupled to the sites $L$ and $R$ in the sample as shown in Fig.
\ref{fig-models}a.

The Floquet Hamiltonian (see Eq.(\ref{eq-Floquet Hamiltonian})) for this
system in the composed space $R\otimes T$ \ is represented in the bottom of
Fig. \ref{fig-models}a. The horizontal dimension corresponds to states
localized in different spatial positions $\left|  i\right\rangle $\ while the
vertical one corresponds to different Fourier states $\left|  n\right\rangle
$. Lines along the horizontal direction correspond to hoppings between states
localized in different positions while the vertical ones are determined by the
Fourier coefficients of the time-dependent part of the Hamiltonian. Thus, we
can clearly see that the resulting picture is that of a higher dimensional
time-independent system. The transport properties can be computed directly
from the Green's functions in this space as discussed before.

Using this picture we can now try to understand the main difference between a
one-parameter and a two-parameter pump.
Symmetry breaking is at the heart of quantum charge pumping: to obtain a
directed current at zero bias potential, the left-right symmetry (LRS) of the
system must be broken. This can be achieved by breaking either time reversal
symmetry (TRS) or inversion symmetry (IS). However, breaking of the LRS alone
does not guarantee a non-vanishing pumped current. In a two terminal
configuration for example even when these symmetries are statically broken
(i.e. through a time-independent potential), TRS under magnetic field
inversion and unitarity of the scattering matrix assures that the
transmittance is insensitive to the direction of propagation
\cite{Buttiker-reciprocity}, i.e. $T_{\rightarrow}(\varepsilon)=T_{\leftarrow
}(\varepsilon)$, and hence there is no pumped current. When a time-dependent
potential is added, photon-assisted processes come into play opening new paths
for transport. The resulting picture is that of a multichannel system and the
reciprocity relation $T_{\rightarrow}(\varepsilon)=T_{\leftarrow}%
(\varepsilon)$ valid for the static case is replaced by an integral relation,
$\int_{0}^{\infty}T_{\rightarrow}(\varepsilon)d\varepsilon=\int_{0}^{\infty
}T_{\leftarrow}(\varepsilon)d\varepsilon$
%
%
%
%
%
%
%
%
%
, thus allowing for a non-vanishing pumping current.
The crucial difference between the situation where only one time-dependent
parameter is present and the one with two, is the possibility of making a
closed loop in Floquet space involving at least two\ vertical processes (or
two ``paths'' in Floquet space). When the two-parameter variation is out of
phase $\varphi_{L}-\varphi_{R}\neq0$, \textit{there will be a non-vanishing
accumulated phase through the loop in a way which is analogous to a magnetic
flux.} For $\varphi_{L}-\varphi_{R}\neq0\operatorname{mod}(\pi)$ the
accumulated phase is different depending on the direction of motion. Note that
this directional asymmetry of the electronic motion (which is a consequence of
a dynamical breaking of IS\ and TRS) is maximum when $\varphi_{L}-\varphi
_{R}=\pi/2\operatorname{mod}(\pi)$.
We will see that for a system as shown in Fig. \ref{fig-schemes}b, where only
one time-dependent gate voltage is present, the directional asymmetry is
provided by the static magnetic field and is manifested as a pumped current
only when photon-assisted processes are allowed.

\section{Model}

In what follows we focus in a system as depicted in Fig. \ref{fig-schemes}b,
consisting of a quantum dot embedded in an arm of a ring which in turn is
connected to two leads (the dot is placed symmetrically between the leads).
The ring is threaded by a magnetic flux $\phi$. We do not consider the
electron-electron interaction in the dot which is a reasonable approximation
for strong dot-ring coupling (open dot). We demonstrate our results using a
lattice Hamiltonian similar to the one used in Ref. \cite{DAmato AB} and more
recently in \cite{Aharony and Imry}. While we use the simplest model with this
geometry, i.e. a four sites ring, our results can be extended to more general
situations involving for example several sites in each arm of the ring and
arbitrary potential profile. Our Hamiltonian is depicted in Fig.
\ref{fig-models}b.

%
%
%
%
%

The magnetic flux $\phi$ is introduced as a phase factor in $V_{L,d}%
=V_{d,L}^{\ast}=\left|  V_{L,d}\right|  \exp(i2\pi\phi/\phi_{0})$ (note that
gauge invariance allows to place it on any bond of the ring), $b$ and $d$ are
the labels used for the site in the reference arm of the ring and the dot
respectively. The energy of the state in the dot $E_{dot}$ which is modulated
periodically by a gate voltage is modeled through $E_{dot}=E_{0}+2V_{g}%
\cos(\omega_{0}t)$. The leads are modeled as one-dimensional tight-binding
chains with zero site energy and hopping matrix element $V$ ($H_{leads}%
=\sum_{\left\langle i,j\right\rangle \text{ with }i,j\in l,r}V(c_{i}^{+}%
c_{j}^{{}}+h.c.)$).

\section{Results and discussion}

Now we turn into the study of the pumping properties of our system. First we
write the kernel in Eq.(\ref{eq-averageCurrent}) as a sum of contributions due
to the different channels:%
\begin{equation}
\sum_{n}\left[  T_{R\leftarrow L}^{(n)}(\varepsilon)-T_{L\leftarrow R}%
^{(n)}(\varepsilon)\right]  \equiv\sum_{n}\delta T_{{}}^{(n)}(\varepsilon
)\equiv\delta T(\varepsilon),
\end{equation}
where%
\begin{align}
\delta T_{{}}^{(n)}(\varepsilon)  &  =T_{R\leftarrow L}^{(n)}(\varepsilon
)-T_{L\leftarrow R}^{(n)}(\varepsilon)\\
&  =2\Gamma_{(R,n)}(\varepsilon)\left|  G_{(R,n)\leftarrow(L,0)}%
^{F}(\varepsilon)\right|  ^{2}\,2\Gamma_{(L,0)}(\varepsilon)-\nonumber\\
&  -2\Gamma_{(L,n)}(\varepsilon)\left|  G_{(L,n)\leftarrow(R,0)}%
^{F}(\varepsilon)\right|  ^{2}\,2\Gamma_{(R,0)}(\varepsilon).\nonumber
\end{align}

To gain an understanding of the physical mechanisms that give rise to pumping
in our system we compute the Floquet-Green's functions up to the first
non-vanishing order in the driving amplitude.
%
%
%
%
%
%
%
%
%
%
%
%
%
%
%
Using Dyson's equation for the Floquet-Green's functions we obtain for the
elastic component:%

\begin{equation}
G_{(R,0)\leftarrow(L,0)}^{F}(\varepsilon)\simeq g_{R,L}^{{}}+g_{R,d}^{{}}%
V_{g}\left(  g_{d,d}^{(+)}+g_{d,d}^{(-)}\right)  V_{g}g_{d,L}^{{}},
\label{eq-approxG_F_elastic}%
\end{equation}
and similarly for $G_{(L,0)\leftarrow(R,0)}^{F}(\varepsilon)$. $g_{i,j}^{{}}$
are the exact retarded Green's functions for the system in the absence of the
time-dependent potential. The superscripts are a short notation to indicate
that the corresponding Green's functions are evaluated at a displaced energy,
$g_{i,j}^{(\pm)}\equiv g_{i,j}^{{}}(\varepsilon\pm\hbar\omega)$, all the other
functions are evaluated at the energy $\varepsilon$.

From Eq. (\ref{eq-approxG_F_elastic}) we can appreciate that the elastic
component of the Floquet-Green's function connecting the left and right
electrodes is the sum of two terms: one that corresponds to direct
transmission from left to right and other that involves virtual photon
emission and absorption in the dot. Thus, the modulus squared of
$G_{(\alpha,0);(\beta,0)}^{F}(\varepsilon)$, contains three terms: \textit{a)}
$|g_{\alpha,\beta}^{{}}|^{2}$, which\ is $\omega$ independent and does not
contribute to the pumped current because it obeys the symmetry $|g_{\alpha
,\beta}^{{}}|=|g_{\beta,\alpha}^{{}}|$ \cite{Buttiker-reciprocity}.
\textit{b)} The next contribution is the modulus squared of the second term in
Eq. (\ref{eq-approxG_F_elastic}) which is fourth order in the driving
amplitude and can be neglected in a first approximation. \textit{c)} The last
contribution is an interference term between the quantum mechanical amplitudes
corresponding to direct tunneling and tunneling plus virtual photon emission
and absorption. This term, which critically depends on the phase difference
between the two terms in Eq. (\ref{eq-approxG_F_elastic}), has a directional
asymmetry due to the presence of the magnetic field which gives rise to a
non-vanishing contribution to the pumped current.

This asymmetry is reflected in the fact that $|g_{L,d}^{{}}|\neq|g_{d,L}^{{}%
}|$ for $\phi\neq0\operatorname{mod}(\pi)$. To understand this difference it
is useful to note that $g_{L,d}^{{}}$ (or $g_{d,L}^{{}}$)\ is proportional to
the effective hopping between the corresponding sites, namely $\widetilde
{V}_{L,d}$ (or $\widetilde{V}_{d,L}$, being the proportionality constant the
same for both cases). This effective hopping can be written as a sum of two
terms, a direct one from site $L$ to the dot and other that corresponds to the
alternative path through the ring: $\widetilde{V}_{L,d}=V_{L,d}+V_{L,b}%
\widetilde{g}_{b}V_{b,R}\overline{g}_{R}V_{R,d}$ (a similar expression holds
for$\,\widetilde{V}_{d,L}$), where $\widetilde{g}_{b}$ is the Green's function
of the isolated $b$ site renormalized by the presence of the $R$ site and the
right lead, $\overline{g}_{R}$ is the Green's function of the isolated right
site renormalized by the right lead. It is easy to see that the interference
between these two spatial paths is directionally asymmetric for $\phi
\neq0\operatorname{mod}(\pi),$ giving therefore $|g_{L,d}^{{}}|$ $\neq
|g_{d,L}^{{}}|$. Again, we have a situation similar to the one in Eq.
(\ref{eq-approxG_F_elastic}) but this time the interference takes place in
real space.

A similar analysis can be performed based on the study of the Floquet-Green's
functions involving a net photon absorption and emission:%
\begin{equation}
G_{(R,\pm1)\leftarrow(L,0)}^{F}(\varepsilon)\simeq g_{R,d}^{(\pm)}V_{g}%
g_{d,L}^{{}}.\label{eq-approxGFinelastic}%
\end{equation}
Again the main observation is that the pumped current is originated from
spatial interference through the ring plus the photon-assisted processes
provided by the time-dependent variation of the dot's energy.

In order to obtain the frequency dependence of the pumped current we assume
the validity of the broad-band approximation and expand the Green's functions
$g_{\alpha\beta}^{(\pm)}=g_{\alpha\beta}^{{}}(\varepsilon\pm\hbar\omega_{0})$
for low frequencies. Using this expansion in Eqs. (\ref{eq-approxG_F_elastic})
and (\ref{eq-approxGFinelastic}) it can be seen that the overall frequency
independent contribution to $\delta T$ is zero and results from a cancellation
between the elastic and the inelastic contributions. The contributions to
$\delta T_{{}}^{(+)}$ that are first order in $\hbar\omega_{0}$ cancel with
the corresponding ones in $\ \delta T_{{}}^{(-)}$. Inspection of Eq.
(\ref{eq-approxG_F_elastic}) shows that the linear term in $\delta T_{{}%
}^{(0)}$ also vanishes as a consequence of the symmetry between the sidebands
of absorption and emission \cite{note-1st-orderContrib}. Hence, we observe
that the first non-vanishing contribution to the average current $\overline
{I}$ is proportional to $\left(  V_{g}\omega_{0}\right)  ^{2}$.

It must be noted that the predicted frequency dependence of the pumped current $\overline
{I}$ is in consistency with the general results presented in Refs. \cite{FalkoSovPhys,VavilovPhaseSpacePicture} for different systems. Here, we interpret our results within the framework introduced in Ref. \cite{VavilovPhaseSpacePicture}: the charge pumped per cycle is determined not by the contour in parameter space as in Ref. \cite{Brouwer PRB1998} (which in this case encloses a vanishing area)  but by the contour in \textit{phase space} \cite{VavilovPhaseSpacePicture}  which contains in addition to the pumping parameters their time-derivatives. The contour in phase space encompasses a non-vanishing area which is proportional to $\omega_{0}$, giving thus the predicted  quadratic frequency dependence for the pumped current $\overline{I}$.

\begin{figure}[ptb]
%
%
%
%
%
\includegraphics[width=8.0cm]{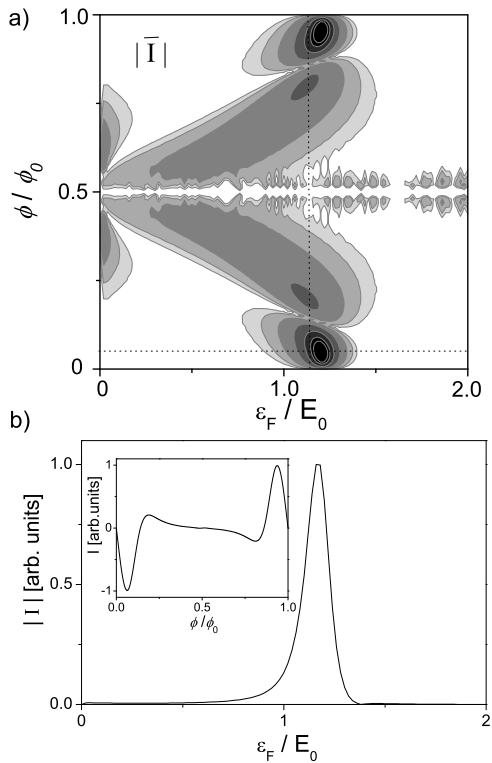} \vspace{0.5cm}\caption{ a) Contour
plot of the the absolute value of the average dc current $\bar{I}$ for the
system shown in \ref{fig-models}b. The horizontal scale corresponds to the
Fermi energy in the leads while the vertical axis is the flux through the
ring. The values of the parameters in this plot are: $V_{g}=0.001$,
$\hbar\omega=0.002$, $\left|  V_{L,d}\right|  =\left|
V_{R,d}\right|  =-0.5 $, all the other hoppings are set equal to $V=-1$ which is
taken as the unit of energy. Note the strong maximum for Fermi energies close
to the dot's energy and flux $\phi\sim0.1\phi_{0}$. In b) we show the
dependence on the Fermi energy for $\phi=0.1\phi_{0}$ (the other parameters
are the same as in a). The inset in b) shows the flux dependence close to the
resonant point ($\varepsilon_{F}=1.15$). The plots in this lower panel correspond to traces along the doted lines in the contour plot.}%
\label{fig-map}%
\end{figure}

\begin{figure}[ptb]
%
%
%
%
%
\includegraphics[width=8.0cm]{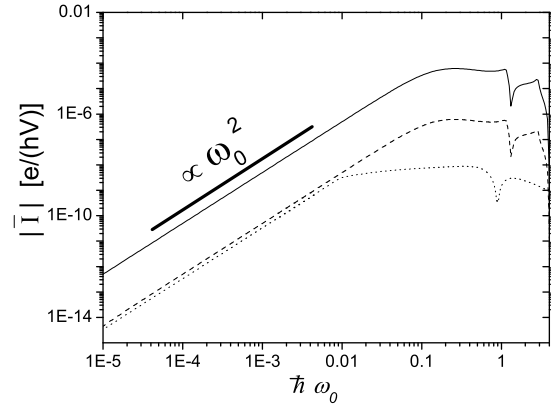} \vspace{0.5cm}\caption{ Absolute value
of the average dc current $\bar{I}$ for the system shown in
\ref{fig-models}b as a function of the driving frequency. The solid and the
dashed lines correspond to Fermi energy and flux close to the resonant point
($\varepsilon_{F}=1.15$, $\phi=0.1\phi_{0}$). \ The difference between both
curves is in magnitude of the driving amplitude, $V_{g}=0.01$ for the solid
line and $V_{g}=0.001$ for the dashed line (the values of the other parameters 
are as in the previous figure). The dotted line corresponds to a small Fermi energy $\varepsilon_{F}=0.01$, $\phi
=0.4\phi_{0}$ and $V_{g}=0.001$. Note that for this case (dotted line) the quadratic dependence with the driving frequency holds up to $\omega_{0}\sim\varepsilon_{F}$.}%
\label{fig-loglog}%
\end{figure}

The results obtained up to now are very general in the sense that they do not
depend much on the specifics of the model as long as the geometry is
preserved. Let us contrast these analytical results with the numerical results
for arbitrary frequency and driving amplitude. The results shown in Figs.
\ref{fig-map} and \ref{fig-loglog} are computed using Eqs.
(\ref{eq-averageCurrent}), (\ref{eq-T_n}) and
(\ref{eq-FourierComponents-FloquetGF}) for zero temperature. Satisfactory
accuracy is obtained by considering Floquet's states with $\left|  n\right|
\leq4$.

A contour plot for the absolute value of the average current for small
frequency and driving amplitudes is shown in Fig. \ref{fig-map}a, the
horizontal axis corresponds to the position of the Fermi energy in the leads
and the vertical one corresponds to the magnetic flux in units of the flux
quantum. The regions with larger $\bar{I}$ values correspond to darker areas.
We observe that the larger currents are located for energies close to the
energy of the dot's level ($\varepsilon_{F}\sim1$) and small magnetic flux
($\phi\sim0.1\,\phi_{0}$). As the magnetic flux increases from zero to half
flux quantum, the position of these maxima move due to interference inside the
ring. At $\phi=0.5\phi_{0}$ the TRS of the system is restored and the pumped
current vanishes.

In Fig. \ref{fig-map}b we show the average current as a function of the Fermi
energy for ($\phi\sim0.1\,\phi_{0}$). We observe a resonant behavior for Fermi
energies close to the dot's energy. This is expected since photon-assisted
processes are stronger close the resonant condition. In the inset of Fig.
\ref{fig-map}b we show the flux dependence near the resonant point
($\varepsilon_{F}=1.15$). The pumped current is periodic in the applied magnetic flux with a period equal to the flux quantum and several harmonics (up to the fifth) contribute importantly to this dependence. Interestingly, we can see that the pumped current can be reversed by tuning
either the magnitude or the direction of the magnetic field.

Other interesting feature that we can appreciate in Fig. \ref{fig-map}a is the
appearance of very narrow maxima in the pumped current for Fermi energies of
the order of the driving frequency (weak maxima close to the vertical
$\varepsilon_{F}=0$ axis in Fig. \ref{fig-map}a ). This is due to the fact
that for energies smaller than $\hbar\omega_{0}$, the processes involving of
photon emission are energetically forbidden thus generating a strong asymmetry
between emission and absorption that leads to a pumped current that decays in
magnitude as the Fermi energy is increased. In this highly non-adiabatic
situation, the previous theoretical analysis based on a low frequency
expansion of the Green's functions fails and the currents do not follow the
predicted $\omega_{0}^{2}$ dependence for $\hbar\omega_{0}\gtrsim
\varepsilon_{F}$ (see dotted line in Fig. \ref{fig-loglog}).

The pumped current as a function of $\hbar\omega_{0}$ is shown in Fig.
\ref{fig-loglog}. Different curves correspond to different values of the
driving amplitude. A line corresponding to a quadratic behavior is also shown
for reference. These results clearly show the predicted low frequency
dependence of the pumped current $\overline{I}\propto\omega_{0}^{2}$ up to
frequencies of the order of the minimum between the width of the resonant
level in the dot ($\Gamma_{dot}\sim$ $0.4$) and the Fermi energy. The
dependence on the driving amplitude (not shown in the figure) also verifies a
quadratic dependence up to moderate driving amplitudes ($V_{g} \sim 0.1V$).

%
%
%
%
%
%
%
%
%
%
%
%
%
%

%
%
%
%
%
%
%
%
%
%
%
%
%
%
%
%
%
%
%
%
%
%
%
%
%
%
%

\section{Conclusions}

In summary, we have studied quantum charge pumping in a system with a
\textit{single} time-periodic parameter using a formalism based on the
computation of the Floquet-Green's functions. The resulting picture is that of
a time-independent system in a higher dimensional space where processes
occurring in real space and photon-assisted processes enter in the same
footing. This allows us to clarify the main differences between a
one-parameter and a two-parameter pump.

Our pump consists of a ring connected to two leads and containing a ``dot''
embedded in one of its arms. The ring is threaded by a magnetic flux while the
dot levels are subjected to a time-periodic gate voltage. We have shown that a
pumped current proportional to the square of the driving frequency appears as
a result of the combined effect of spatial interference through the ring and
photon-assisted tunneling. The direction of the current can be changed by
tuning either the direction or the magnitude of the magnetic field. It must be
emphasized that the directional asymmetry needed to obtain quantum pumping is
provided through the static magnetic field while the unique role of the
time-dependent parameter is to provide additional inelastic channels for
transport. In this sense, a pumped current can be obtained using any other
mechanism that provides such inelastic channels as long as the phase coherence
of the composed system (sample plus inelastic scatterer) is preserved.

\section{Acknowledgments}

The author acknowledges V. E. Kravtsov for useful discussions and H. M.
Pastawski for helpful comments. The author is also indebted to Saugata Ghosh
for careful reading of the manuscript.

\section{{Appendix}}

In order to derive Eq. (\ref{eq-FourierComponents-FloquetGF}), we write the
retarded Green's function $G_{\beta,\alpha}(t,\varepsilon)$ in terms of the
time-evolution operator of the system $U_{\beta,\alpha}(t,t^{\prime})$
(defined by the relations $\left|  \psi(t)\right\rangle =U(t,t_{0})\left|
\psi(t_{0})\right\rangle ,\,\,\,\,U(t_{0},t_{0})\equiv1$):%
\begin{equation}
G_{\beta,\alpha}(t,\varepsilon)=-\frac{i}{\hbar}\int_{0}^{\infty}d\tau
\exp(i\varepsilon\tau/\hbar)U_{\beta,\alpha}(t,t-\tau). \label{eq-Appendix}%
\end{equation}
Then, using the well known relation between the matrix elements of the
time-evolution operator and the Floquet Hamiltonian\cite{Shirley}%

\begin{align}
\left\langle \beta\right|  U(t,t_{0})\left|  \alpha\right\rangle  &
=\sum_{n=-\infty}^{\infty}\left\langle \beta,n\right|  \exp(-iH_{F}%
(t,t_{0})/\hbar)\left|  \alpha,0\right\rangle \\
&  \times\exp(in\omega t)\nonumber
\end{align}
in Eq. (\ref{eq-Appendix}) and integrating over $\tau$ we obtain,%
\begin{equation}
G_{\beta,\alpha}(t,\varepsilon)=\sum_{n=-\infty}^{\infty}\left\langle
\beta,n\right|  (\varepsilon1-H_{F})^{-1}\left|  \alpha,0\right\rangle
\exp(in\omega t).
\end{equation}
The coefficients of the exponential can be identified as the Fourier
coefficients in the Fourier expansion of $G_{\beta,\alpha}(t,\varepsilon
)=\sum_{n}G_{\beta,\alpha}^{(n)}(\varepsilon)\exp(in\omega t)$ from where Eq.
(\ref{eq-FourierComponents-FloquetGF}) follows.

Substituting this relation into Eqs.(\ref{eq-T_n}) and
(\ref{eq-averageCurrent}) gives the average current in terms of the
Floquet-Green's functions:%
\begin{equation}
\overline{I}=\frac{e}{h}\sum_{n}\int d\varepsilon\left[  T_{R\leftarrow
L}^{(n)}(\varepsilon)-T_{L\leftarrow R}^{(n)}(\varepsilon)\right]
f(\varepsilon)
\end{equation}
where
\[
T_{R\leftarrow L}^{(n)}(\varepsilon)=2\Gamma_{(R,n)}(\varepsilon)\left|
G_{(R,n)\leftarrow(L,0)}^{F}(\varepsilon)\right|  ^{2}\,2\Gamma_{(L,0)}%
(\varepsilon).
\]
The transmittance in the reverse sense follows from the last equation by
exchanging the $L$ and $R$ indexes.


\begin{thebibliography}{99}
\bibitem{Intro-Floquet} For an introduction to Floquet theory we refer the reader to Refs. \cite{Shirley,Sambe}. For a more recent review see \cite{Kohler PhysRep}.
\bibitem{note-1st-orderContrib}Note that for the system shown in Fig.
\ref{fig-models}a there is a directional asymmetry in the photon emission and
absorption processes whenever $\varphi\neq0\operatorname{mod}(\pi)$. This
produces a non-vanishing linear in $\omega$ contribution to the pumped current.
\end{thebibliography}
\end{document}